\documentstyle[12pt]{article}

\expandafter\ifx\csname mathrm\endcsname\relax\def\mathrm#1{{\rm #1}}\fi

\makeatletter
\if@twoside \oddsidemargin -17pt \evensidemargin 00pt
\else \oddsidemargin 00pt \evensidemargin 00pt
\fi
\expandafter\ifx\csname mathrm\endcsname\relax\def\mathrm#1{{\rm #1}}\fi

\makeatother

\def\beq{\begin{equation}}
\def\eeq{\end{equation}}
\def\beqar{\begin{eqnarray}}
\def\eeqar{\end{eqnarray}}
\def\barr#1{\begin{array}{#1}}
\def\earr{\end{array}}
\def\bfi{\begin{figure}}
\def\efi{\end{figure}}
\def\btab{\begin{table}}
\def\etab{\end{table}}
\def\bce{\begin{center}}
\def\ece{\end{center}}
\def\nn{\nonumber}

\def\text{\textstyle}

\def\reffi#1{\mbox{Fig.~\ref{#1}}}
\def\reffis#1{\mbox{Figs.~\ref{#1}}}
\def\refta#1{\mbox{Tab.~\ref{#1}}}
\def\refse#1{\mbox{Sect.~\ref{#1}}}

\def\refapp#1{\mbox{App.~\ref{#1}}}
\def\citere#1{\mbox{Ref.~\cite{#1}}}

\newcommand{\GeV}{\unskip\,\mathrm{GeV}}
\newcommand{\MeV}{\unskip\,\mathrm{MeV}}

\renewcommand{\O}{{\cal O}}

\def\mathswitchr#1{\relax\ifmmode{\mathrm{#1}}\else$\mathrm{#1}$\fi}

\newcommand{\PW}{\mathswitchr W}
\newcommand{\PZ}{\mathswitchr Z}

\newcommand{\PH}{\mathswitchr H}
\newcommand{\Pe}{\mathswitchr e}

\newcommand{\Pane}{\mathswitch \bar\nu_{\mathrm{e}}}
\newcommand{\Pnmu}{\mathswitch \nu_\mu}

\newcommand{\Pt}{\mathswitchr t}

\newcommand{\Pem}{\mathswitchr {e^-}}
\newcommand{\Pmum}{\mathswitchr {\mu^-}}

\newcommand{\PWpm}{\mathswitchr {W^\pm}}
\newcommand{\PWO}{\mathswitchr {W^0}}

\def\mathswitch#1{\relax\ifmmode#1\else$#1$\fi}

\newcommand{\MW}{\mathswitch {M_\PW}}
\newcommand{\MWpm}{\mathswitch {M_\PWpm}}
\newcommand{\MWO}{\mathswitch {M_\PWO}}

\newcommand{\MZ}{\mathswitch {M_\PZ}}
\newcommand{\MH}{\mathswitch {M_\PH}}

\newcommand{\Mt}{\mathswitch {m_\Pt}}

\newcommand{\sw}{\mathswitch {s_\PW}}

\newcommand{\logx}{f_1}
\newcommand{\logy}{f_2}
\newcommand{\alpz}{\alpha(\MZ^2)}
\newcommand{\bos}{{\mathrm{bos}}}
\newcommand{\fer}{{\mathrm{ferm}}}
\newcommand{\uni}{{\mathrm{univ}}}
\newcommand{\BW} {{\mathrm{WPD}}}
\newcommand{\BZ} {{\mathrm{ZPD}}}

\def\atn{\mathop{\mathrm{arctan}}\nolimits}
\def\Li{\mathop{\mathrm{Li}_2}\nolimits}
\def\Re{\mathop{\mathrm{Re}}\nolimits}

\def\solid{\raise.9mm\hbox{\protect\rule{1.1cm}{.2mm}}}

\def\draftdate{\relax}
\def\mda{\relax}
\def\mua{\relax}
\def\mla{\relax}
\def\mpar#1{\relax}
\def\draft{
\def\draftdate{\today}
\def\mpar##1{\marginpar{\hbadness10000\sloppy\boldmath\bf##1}%
                      \typeout{marginpar: \noexpand##1}\ignorespaces}
\def\mda{\mpar{\hfil$\downarrow$\hfil}}
\def\mua{\mpar{\hfil$\uparrow$\hfil}}
\def\mla{\marginpar[\boldmath\hfil$\rightarrow$\hfil]%
                   {\boldmath\hfil$\leftarrow $\hfil}%
                    \typeout{marginpar: $\leftrightarrow$}\ignorespaces}
}

\makeatletter

\def\eqnarray{\stepcounter{equation}\let\@currentlabel=\theequation
\global\@eqnswtrue
\global\@eqcnt\z@\tabskip\@centering\let\\=\@eqncr
$$\halign to \displaywidth\bgroup\hskip\@centering
  $\displaystyle\tabskip\z@{##}$\@eqnsel&\global\@eqcnt\@ne
  \hskip 2\arraycolsep \hfil${##}$\hfil
  &\global\@eqcnt\tw@ \hskip 2\arraycolsep $\displaystyle\tabskip\z@{##}$\hfil
   \tabskip\@centering&\llap{##}\tabskip\z@\cr}
\def\appendix{\par
 \setcounter{section}{0} \setcounter{subsection}{0}
 \def\thesection{\Alph{section}}}

\makeatother

\newcommand{\lsim}
{\;\raisebox{-.3em}{$\stackrel{\displaystyle <}{\sim}$}\;}

\begin{document}
\thispagestyle{empty}
\def\thefootnote{\fnsymbol{footnote}}
\setcounter{footnote}{1}
\null
\hfill BI-TP 94/09 \\
\null
\vskip .8cm
\begin{center}
{\Large \bf On the Significance of the Electroweak Precision Data \par}
\vskip 3em
{\large S.\ Dittmaier%
\footnote{Supported by the Bundesminister f\"ur Forschung
und Technologie, Bonn, Germany.}%
, D.\ Schildknecht}
\vskip .5em
{\it Fakult\"at f\"ur Physik, Universit\"at Bielefeld, Germany}
\vskip 2em
{\large K.\ Kolodziej}
\vskip .5em
{\it Theoretische Physik, Universit\"at M\"unchen, Germany}
\vskip 2em
{\large M.\ Kuroda}
\vskip .5em
{\it Department of Physics, Meiji-Gakuin University, Yokohama, Japan}
\end{center} \par
\vskip 1.2cm
\vfil
{\bf Abstract} \par
We elaborate on a recently suggested effective Lagrangian for
charged-current and neutral-current electroweak interactions which
in comparison with the standard electroweak theory contains three
free parameters $\Delta x, \Delta y$, $\varepsilon$ which quantify
different sources for violations of $SU(2)$ symmetry. Within the
standard $SU(2)_I \times U(1)_Y$ electroweak theory, we present both
exact and very much refined approximate analytical one-loop
expressions for these parameters in terms of the canonical
input, $G_\mu, \MZ$, $\alpz$, the top-quark mass, $\Mt$, and the
Higgs-boson mass, $\MH$.
We reemphase the importance of discriminating between the
{\it empirically well-known purely fermionic} (vacuum polarization)
contributions to $\Delta x, \Delta y$, $\varepsilon$ and the
{\it empirically unknown bosonic} ones with respect to present and future 
electroweak precision tests. The parameters $\Delta x$ and $\varepsilon$
are hardly affected by standard bosonic corrections, while the full
one-loop results for $\Delta y$ differ appreciably from the ones
obtained by taking into account fermion loops only.
A detailed  comparison with the experimental data on $\MWpm/\MZ$, $\bar\sw^2$,
$\Gamma_l$ shows that these data start to become accurate enough to be
sensitive to standard (bosonic) contributions to $\Delta y$ beyond 
fermion loops.
\par
\vskip 1cm
\noindent March 1994 \par
\null
\setcounter{page}{0}
\clearpage
\def\thefootnote{\arabic{footnote}}
\setcounter{footnote}{0}

\section{Introduction}
\label{intro}

We have recently given \cite{bi93} an effective Lagrangian for
neutral- and charged-current electroweak vector-boson fermion 
interactions which contains three additional free parameters in 
comparison with the standard electroweak
$SU(2)_I \times U(1)_Y$ gauge theory. The three parameters, called
$\Delta x, \Delta y$, $\varepsilon$, specify the strengths of
various $SU(2)$-violating terms in the underlying Lagrangian. They are
related to the charged and neutral boson masses, the charged-current and
neutral-current couplings and a kinetic mixing term in the
neutral-current sector, respectively. The purpose for analysing this
effective Langrangian is (at least) twofold \cite{bi93}:
\renewcommand{\labelenumi}{(\roman{enumi})}
\begin{enumerate}
\item The effective Lagrangian, when evaluated at tree level,
leads to predictions for the observables%
\footnote{We use the standard notation, $\MWpm$ and $\MZ$, for the
masses of charged and neutral weak bosons, respectively, $\bar\sw^2$ for the
square of the weak leptonic mixing angle measured at $q^2=\MZ^2$ and
$\Gamma_l$ for the leptonic decay width of the Z boson, etc. In the
present paper we confine
our analysis to the mentioned observables, as these are particularly
simple ones which do not involve important hadronic (gluonic)
effects.}
$\MWpm/\MZ$, $\bar\sw^2$, $\Gamma_l$
which differ from the standard ones due to the
presence of the additional parameters $\Delta x, \Delta y$,
and $\varepsilon$. Comparison with experiment leads to constraints
in $\Delta x, \Delta y, \varepsilon$, which allow one to quantify
to what extent the fundamental $SU(2)_I \times U(1)_Y$ symmetry
of the standard electroweak theory in the vector-boson fermion
sector is actually valid in nature.
\item Within the standard electroweak theory the parameters
$\Delta x, \Delta y$, $\varepsilon$ can be used to specify the
one-loop radiative corrections in terms of the empirically known and
unknown input parameters $G_\mu$, $\alpz$, $\MZ$ and $\Mt$, $\MH$,
respectively. The parameters $\Delta x, \Delta y$, $\varepsilon$
are well suited to implement the 1988 suggestion
\cite{go88} of Gounaris and one of the present authors of clearly
discriminating the one-loop effects originating from the {\it empirically
well-known} interactions of the vector bosons with (light) fermions and
the {\it empirically unknown} interactions of the bosons among each other.
Evaluating the loop corrections to $\Delta x$, $\Delta y$, $\varepsilon$
by taking into account fermion loops only and comparing with the full
one-loop results allows one to set the scale \cite{go88} for the accuracy
which has to be reached in precision experiments if the standard theory is
to be quantitatively tested beyond fermion loops. In other words,
by comparing both the fermion-loop predictions as well as the full
one-loop predictions with the electroweak precision data for
$\Delta x$, $\Delta y$, $\varepsilon$ (obtained from $\MWpm/\MZ$, $\bar\sw^2$,
and the leptonic width, $\Gamma_l$) allows one to judge in how far
precision data are able to {\it isolate} and thus to {\it directly measure}
standard (bosonic) contributions beyond the empirically
well-known fermion loops. More generally,
comparison of the dominant-fermion-loop predictions with the
data quantifies the extent to which additional effects, standard or
non-standard ones, are required by precision data.
\end{enumerate}

The present paper expands on previous work \cite{bi93,kn91}.
We explicitly give the complete analytic one-loop expressions
for $\Delta x$, $\Delta y$, $\varepsilon$ in the standard
$SU(2)_I\times U(1)_Y$ electroweak theory and explicitly compare the
fermion-loop contributions with the full one-loop results. It turns
out that the parameters $\Delta x$, $\Delta y$, $\varepsilon$ are
particularly well-suited parameters for identifying contributions beyond
fermion loops. While $\varepsilon$ and $\Delta x$ are so strongly
dominated by fermion-loop contributions that experiments will
hardly ever be able to discriminate fermion loops from standard
full one-loop results, the standard bosonic contributions to $\Delta y$, in
contrast, turn out to be quite large. In fact, with the present data we are
starting to be able to discriminate $\Delta y$ as calculated from
fermion loops alone from $\Delta y$ as obtained by taking into account
bosonic self-couplings as well. In this respect, it is of interest that
$\Delta y$ is the only one of the three parameters which contains
the process specific vertex and box corrections to muon decay which
enter the analysis of radiative effects via the input parameter $G_\mu$.
Accordingly, we conclude that the neutral-current
precision data at the Z resonance start to quantitatively test this 
important charged-current one-loop correction to muon decay.

In \refse{lagrange}, we briefly rederive our effective Lagrangian.
In \refse{xyeps}, we present exact and approximate explicit one-loop
formulae for $\Delta x$, $\Delta y$, $\varepsilon$ and
$\MWpm/\MZ$, $\bar\sw^2$, $\Gamma_l$ within the standard
electroweak theory. Our approximate expressions for $\Delta x$,
$\Delta y$, $\varepsilon$, which are based on asymptotic expansions for
$\Mt,\MH\to\infty$, have a transparent and
compact analytic form and are sufficiently accurate for all practical
purposes. The analytic expressions for the remainder terms are
collected in \refapp{xyerem}, while \refapp{aux} lists a
few simple auxiliary functions. In \refse{discuss}, we compare
the most recent empirical results for $\MWpm/\MZ$, $\bar\sw^2$, $\Gamma_l$
with the full one-loop and the fermion-loop predictions.
Final conclusions are drawn in \refse{concl}.

\section{The effective Lagrangian}
\label{lagrange}

Introducing a charged $\PWpm$ vector-boson of mass $\MW$ and coupling
$g_\PWpm$ to the weak isospin current, $j^\pm_\mu$, we have for
charged-current interactions
\beq
{\cal L}_C = -\frac{1}{2} W^{+\mu\nu}W^-_{\mu\nu}
+ \frac{g_\PWpm}{\sqrt 2} \left( j^+_\mu W^{+\mu} + h.c.\right)
+ \MWpm^2 W^+_\mu W^{-\mu}.
\label{cclag}
\eeq
In the transition to the neutral-current sector, we allow for
breaking of $SU(2)$ symmetry in the coupling via the parameter $y$,
\beq
g^2_\PWpm = y g_\PWO^2 = (1 + \Delta y) g^2_\PWO,
\label{dy}
\eeq
and in the mass via the parameter $x$,
\beq
\MWpm^2 = x \MWO^2 = (1 + \Delta x) \MWO^2.
\label{dx}
\eeq
In addition, we introduce current mixing of strength $\lambda$ with
the photon field%
\footnote{Equivalently, one may work in the $W^3B$ base \cite{kn91}.}
\`a la Hung-Sakurai \cite{hu78}
\beq
{\cal L}_{mix} = \lambda A_{\mu\nu} W^{3\mu\nu}.
\label{mixlag}
\eeq
Upon diagonalisation, the neutral-current Lagrangian in the physical 
base reads
\beqar
{\cal L}_N & = & - \frac{1}{4} Z_{\mu\nu} Z^{\mu\nu} +
\frac{1}{2} \frac{\MWO^2}{1-\lambda^2} Z_\mu Z^\mu
- \frac{1}{4} A_{\mu\nu} A^{\mu\nu} \nn\\
&& - e j_{em}^\mu A_\mu - \frac{g_\PWO}{\sqrt{1-\lambda^2}}
\left( j^3 - \lambda\frac{e}{g_\PWO} j_{em}\right)^\mu Z_\mu.
\label{nclag1}
\eeqar
The electromagnetic coupling at the $\PZ$ mass has been denoted by $e$
with
\beq
\frac{e^2(\MZ^2)}{4\pi} = \alpz \approx 1/129,
\eeq
where $\alpz$ includes the `running' of electromagnetic vacuum
polarization induced by the light fermions \cite{alpz}.
Upon introducing the empirical weak angle at the $\PZ$ mass,
$\bar\sw^2$, via
\beq
\bar\sw^2 \equiv \lambda\frac{e}{g_\PWO},
\eeq
and replacing $\lambda$ by $\varepsilon$ via
\beq
\lambda \equiv \frac{e}{g_\PWO} (1-\varepsilon),
\label{lameps}
\eeq
we have
\beq
\bar\sw^2 = \frac{e^2}{g_\PWO^2} (1-\varepsilon).
\label{swrel}
\eeq
The neutral-current interaction Lagrangian (\ref{nclag1}) then
becomes \cite{bi93}
\beqar
{\cal L}_N & = & - \frac{1}{4} Z_{\mu\nu} Z^{\mu\nu} +
\frac{1}{2} \frac{\MWO^2}{1-\bar\sw^2(1-\varepsilon)}Z_{\mu}Z^{\mu}
- \frac{1}{4} A_{\mu\nu} A^{\mu\nu} \nn\\
&& - e j_{em}^\mu A_\mu -
\frac{g_\PWO}{\sqrt{1-\bar\sw^2 (1-\varepsilon)}}
\left( j^\mu_3 - \bar\sw^2 j^\mu_{em}\right) Z_\mu.
\label{nclag2}
\eeqar
In total, we have introduced two parameters in the charged-current
sector, $\MWpm$, $g_\PWpm$, and four parameters related to the
neutral-current sector, $\MWO$ (or, equivalently $\Delta x$ in
(\ref{dx})), $g_\PWO$ (or, equivalently $\Delta y$ in (\ref{dy})),
$e(\MZ^2)$, and $\lambda$ (or equivalently $\varepsilon$ in
(\ref{lameps})). The weak mixing angle $\bar\sw^2$ is related
to these parameters according to (\ref{swrel}).

The standard electroweak interaction Lagrangian \cite{gsw} for
vector-boson fermion interactions is contained in (\ref{nclag2}) in the
limit of $\varepsilon, \Delta x, \Delta y \to 0$.
This is also evident from the expressions for the observable parameters
$\MWpm/\MZ$, $\bar\sw^2$, and the leptonic Z width, $\Gamma_l$, in terms of the
canonical input, the Fermi coupling measured in muon decay
\beq
\frac{8G_\mu}{\sqrt 2} = \frac{g^2_\PWpm}{\MWpm^2},
\label{gmu}
\eeq
the $\PZ$ mass, $\MZ$, and the electromagnetic coupling, $\alpz$, at the
Z resonance.

From the Lagrangian one easily obtains
\beqar
\bar\sw^2 \left( 1-\bar\sw^2 \right)
& = &
\frac{\pi\alpz} {\sqrt 2 G_\mu \MZ^2}
\frac {y}{x} (1-\varepsilon)
\frac{1} {\left( 1 +\frac{\bar\sw^2} {1-\bar\sw^2}
\varepsilon \right)}, \nn\\
\frac {\MWpm^2} {\MZ^2}
& = &
\left( 1-\bar\sw^2 \right)x
\left( 1 + \frac{\bar\sw^2}{1-\bar\sw^2}\varepsilon \right), \nn\\
\Gamma_l
& = &
\frac {G_\mu \MZ^3} {24\pi \sqrt 2}
\left( 1 + \left( 1 - 4\bar\sw^2\right)^2\right)
\frac{x}{y}
\left( 1 - \frac{3\alpha}{4\pi}\right).
\label{obs}
\eeqar
A QED correction factor is included in $\Gamma_l$ in agreement with
the definition of $\Gamma_l$ used in the analysis of experimental
data. Omitting this factor and taking the limit $\Delta x, \Delta y,
\varepsilon \to 0$ indeed yields the well-known standard tree-level
relations.

A comparison of (\ref{obs}) with experiment clearly allows to constrain
$\Delta x, \Delta y$, $\varepsilon$, and accordingly allows
one to quantify the extent to which the standard $SU(2)_I \times U(1)_Y$
symmetry is supported by empirical data.

Alternatively, the Lagrangians (\ref{cclag}) and (\ref{nclag2}) may be
interpreted as
one-loop effective Lagrangians of the standard electroweak theory.
As previously noted, the standard radiative corrections to
$\MWpm/\MZ$, $\bar\sw^2$, $\Gamma_l$ may indeed
be incorporated by appropriate specification of the $SU(2)$-breaking
parameters, $\Delta x$, $\Delta y$, $\varepsilon$. The explicit
expressions and very accurate approximations for $\Delta x$, $\Delta y$,
$\varepsilon$ in terms of the known input $G_\mu$, $\alpz$
and $\MZ$ and the unknown parameters $\Mt$ and $\MH$ are given
in the \refse{xyeps}.

We close this section by noting that our parameters $\Delta x$,
$\Delta y$, $\varepsilon$, introduced and defined in terms of
$SU(2)$-symmetry properties of electroweak interactions are simple
linear combinations of the parameters $\varepsilon_{Ni}$ $(i=1,2,3)$
introduced in \citere{al93} by the requirement of isolating the quadratic
$\Mt$-dependence in the one-loop standard formulae for these parameters,
\beq
\Delta x =  \varepsilon_{N1}-\varepsilon_{N2}, \quad
\Delta y = -\varepsilon_{N2}, \quad
\varepsilon = -\varepsilon_{N3}.
\label{alteps}
\eeq
These relations between the $\varepsilon_{Ni}$ and our parameters
establish the meaning of the $\varepsilon_{Ni}$ with respect to
$SU(2)$-symmetry properties of electroweak interactions.

\section{Analytic results for the parameters $\Delta x$, $\Delta y$,
$\varepsilon$}
\label{xyeps}

In this section, we present the results for $\Delta x$, $\Delta y$,
$\varepsilon$ within the electroweak standard model which are obtained
by analyzing the decays $\Pmum\to\Pnmu\Pem\Pane$ and $\PZ\to l\bar l$
$(l=e,\mu,\tau)$. More precisely, one expands (\ref{obs}) in linear order
in $\Delta x$, $\Delta y$, $\varepsilon$ to obtain \cite{bi93}
\beqar
\bar\sw^2 &=& s_0^2\left[1-\frac{1}{c_0^2-s_0^2}\varepsilon
-\frac{c_0^2}{c_0^2-s_0^2}(\Delta x-\Delta y)\right], \nn\\
\frac{\MWpm}{\MZ} &=& c_0\left[1+\frac{s_0^2}{c_0^2-s_0^2}\varepsilon
+\frac{c_0^2}{2(c_0^2-s_0^2)}\Delta x
-\frac{s_0^2}{2(c_0^2-s_0^2)}\Delta y
\right], \nn\\
\Gamma_l &=& \Gamma_l^{(0)}\left[1
+\frac{8s_0^2(1-4s_0^2)}{(c_0^2-s_0^2)(1+(1-4s_0^2)^2)}\varepsilon
+\frac{2(c_0^2-s_0^2-4s_0^4)}{(c_0^2-s_0^2)(1+(1-4s_0^2)^2)}
(\Delta x-\Delta y)\right], \hspace*{1.9em}
\label{obslin}
\eeqar
where
\beqar
&& s_0^2(1-s_0^2)\equiv c_0^2s_0^2 = \frac{\pi\alpz}{\sqrt{2}G_\mu\MZ^2},
\nn\\
\Gamma_l^{(0)} &=&  \frac{\alpz\MZ}{48s_0^2c_0^2}
\left[1+(1-4s_0^2)^2\right]\left(1+\frac{3\alpha}{4\pi}\right),
\eeqar
and identifies $\Delta x$, $\Delta y$, $\varepsilon$ by a comparison with
the electroweak radiative corrections to 
$\MWpm/\MZ$, $\bar\sw^2$, $\Gamma_l$ within the standard model. For the
explicit calculations of the corresponding one-loop radiative corrections
we have applied well-known standard techniques, which are e.g.\ summarized
in \citere{admex}.

The contributions to $\Delta x$, $\Delta y$, $\varepsilon$ are
decomposed into gauge invariant subsets as far as possible. As already
emphasized, we
separate those contributions from the bulk of the radiative corrections
which originate from Feynman graphs containing a closed fermion loop.
Obviously this (gauge invariant) part is universal -- as it enters
merely via vacuum polarization -- and is denoted by $a_\fer^\uni$
$(a = \Delta x, \Delta y, \varepsilon)$. The remaining part of the 
radiative corrections is split into a bosonic universal contribution, 
$a_\bos^\uni$, and
two process dependent ones, $a^\BW$ and $a^\BZ$, which are related
to the muon (WPD) and the Z (ZPD) decay, respectively. Accordingly, we
have the decomposition
\beq
a = a_\fer^\uni + a_\bos^\uni + a^\BW + a^\BZ, \qquad\mbox{with}\quad
a = \Delta x, \Delta y, \varepsilon.
\label{xyesplit}
\eeq
Note that the (gauge invariant) decomposition of bosonic radiative corrections
into process dependent and independent parts is not uniquely determined
by the criterion of gauge invariance so that supplementary conditions
have to be imposed. Various suggestions and methods on this subject
were accordingly given in the literature by different authors
\cite{gauginv1,gauginv2,pt}. Here we decided to apply the so-called
`pinch technique' \cite{pt} which provides a technically particularly simple
prescription for the decomposition of transition amplitudes into
gauge invariant self-energy, vertex, and box contributions.

We split the universal parts,
$a_{\fer/\bos}^\uni$, further into `dominant' parts,
$a_{\fer/\bos}^\uni (dom)$, and remainders, $a_{\fer/\bos}^\uni (rem)$,
\beq
a_{\fer/\bos}^\uni = a_{\fer/\bos}^\uni (dom) + a_{\fer/\bos}^\uni(rem),
\label{domrem}
\eeq
where $a_{\fer/\bos}^\uni (dom)$ is defined as the asymptotic expansion of the
full universal contribution, $a_{\fer/\bos}^\uni$, in the limit of a heavy
top quark and a heavy Higgs boson, i.e.\ for $t\to\infty$, $h\to\infty$ with
\beq
t=\frac{\Mt^2}{\MZ^2}, \qquad h=\frac{\MH^2}{\MZ^2},
\eeq
keeping terms up to constant order. As will be discussed below these
asymptotic formulae approximate $a_{\fer/\bos}^\uni$ very well in the
relevant regions for $t$ and $h$, so that the terminology `dominant'
is justified. In this context we mention that the remaining (light)
fermions, i.e.\ all leptons and the light quarks, are taken exactly
massless since $\alpz$ already includes all sizeable effects arising
from light fermion masses.

The explicit expressions for the fermionic contributions read
\beqar
\Delta x_\fer^\uni (dom) &=&
\frac{\alpz}{8\pi s_0^2}\left[
\frac{3}{2c_0^2}t+\log(t)+6\log(c_0^2)
-\frac{197}{18}+\frac{103}{9c_0^2}-\frac{40s_0^2}{9}-\frac{160s_0^4}{9}
\right],\nn\\[.3em]
\Delta y_\fer^\uni (dom) &=&
\frac{\alpz}{8\pi s_0^2}\left[
\log(t)+6\log(c_0^2)
-\frac{13}{2}+\frac{40s_0^2}{3}-\frac{160s_0^4}{9}
\right],\nn\\[.3em]
\varepsilon_\fer^\uni (dom) &=&
\frac{\alpz}{24\pi s_0^2}\left[
\log(t)-\frac{125}{6}+40s_0^2-\frac{160s_0^4}{3}
\right],
\label{xyeferdom}
\eeqar
which of course include the well-known $\Mt^2$ \cite{ve77} and 
$\log\Mt$ terms
already given in \citere{bi93}. An analogous statement holds for the
$\log\MH$ terms of the universal bosonic contributions
\beqar
\Delta x_\bos^\uni (dom) &=&
\frac{\alpz}{16\pi s_0^2}\left[
-\frac{3s_0^2}{c_0^2}\log(h)
-\frac{s_0^2}{c_0^4}\left(\frac{233}{6}-\frac{1591s_0^2}{6}
+500s_0^4-\frac{1060s_0^6}{3}+80s_0^8\right)
\right.\nn\\ &&\phantom{\frac{\alpz}{24\pi s_0^2}}\left.
+\left(\frac{23}{2}-\frac{68s_0^2}{3}+8s_0^4+3s_0^6\right)
\frac{\log(c_0^2)}{c_0^6}
-(3-4s_0^2)
\right.\nn\\ &&\phantom{\frac{\alpz}{24\pi s_0^2}}\left.
\times(33-44s_0^2+12s_0^4)\frac{\logy(c_0^2)}{6c_0^6}
-(33+22s_0^2-100s_0^4+40s_0^6)\logx(c_0^2)
\right], \nn\\[.3em]
\Delta y_\bos^\uni (dom) &=&
\frac{\alpz}{16\pi s_0^2}\left[
\frac{1}{c_0^4}\left(40-\frac{1363s_0^2}{6}+\frac{1139s_0^4}{2}
-704s_0^6+\frac{1204s_0^8}{3}-80s_0^{10}\right)
\right.\nn\\ &&\phantom{\frac{\alpz}{24\pi s_0^2}}\left.
-\left(1-\frac{19s_0^2}{2}-\frac{32s_0^4}{3}+6s_0^6-8s_0^8\right)
\frac{\log(c_0^2)}{s_0^2c_0^6}
-(3-4s_0^2)
\right.\nn\\ &&\phantom{\frac{\alpz}{24\pi s_0^2}}\left.
\times(33-44s_0^2+12s_0^4)\frac{\logy(c_0^2)}{6c_0^6}
-(7+78s_0^2-124s_0^4+40s_0^6)\logx(c_0^2)
\right], \nn\\[.3em]
\varepsilon_\bos^\uni (dom) &=&
\frac{\alpz}{48\pi s_0^2}\left[
-\log(\frac{h}{c_0^2})
+\frac{1063}{6}-760s_0^2+820s_0^4-240s_0^6
\right.\nn\\ &&\phantom{\frac{\alpz}{24\pi s_0^2}}\left.
+(105-406s_0^2+420s_0^4-120s_0^6)\logx(c_0^2)
\right].
\label{xyebuni}
\eeqar
The formulae for $a_{\fer/\bos}^\uni (rem)$ and the auxiliary functions
$\logx(x)$, $\logy(x)$ appearing in (\ref{xyebuni}) are listed in
\refapp{xyerem} and \ref{aux}, respectively.
According to (\ref{xyebuni}), the dominant universal bosonic contributions
$\Delta x_\bos^\uni (dom)$ and $\varepsilon_\bos^\uni (dom)$ depend
on the Higgs mass via $\log(h)$, while $\Delta y_\bos^\uni (dom)$
is a constant in the sense of being independent of $h$.

The process dependent parts for the Z decay are obviously independent of
the top-quark and Higgs-boson masses and are given by
\beqar
\Delta x^\BZ &=&
-\frac{\alpz s_0^4}{4\pi c_0^2}\left[ 11+16C_1 \right], \nn\\[.3em]
\varepsilon^\BZ &=&
\frac{\alpz}{4\pi s_0^2}\left[
(1-6s_0^2+16s_0^4-8s_0^6)\frac{2C_1}{c_0^2}
-(2-s_0^2)^2C_2-2c_0^4(3-s_0^2)C_3
\right.\nn\\ &&\phantom{\frac{\alpz}{24\pi s_0^2}}\left.
-\left(\frac{5}{2}-s_0^2\right)
\left(\log(c_0^2)-2c_0^2\logx(c_0^2)\right)
+\frac{17}{8c_0^2}-\frac{27s_0^2}{2c_0^2}+\frac{57s_0^4}{2c_0^2}
-\frac{13s_0^6}{c_0^2}
\right],
\nn\\[.3em]
\Delta y^\BZ &=& \frac{c_0^2}{s_0^2}\Delta x^\BZ+2\varepsilon^\BZ.
\label{dxyepd}
\eeqar
The explicit formulae for the constants $C_1$, $C_2$, $C_3$ can be found
in \refapp{aux}. As already mentioned, only $\Delta y$ gets process
dependent contributions induced by vertex and box corrections to muon decay,
\beq
\Delta y^\BW = \frac{\alpz}{8\pi s_0^4}\left[
4s_0^4+(4s_0^4-1)\log(c_0^2) \right],
\label{ywpd}
\eeq
so that $\Delta x^\BW$ and $\varepsilon^\BW$ are identically zero,
\beq
\Delta x^\BW = \varepsilon^\BW = 0. \nn\\[.3em]
\label{xewpd}
\eeq
Recall that all genuine vertex corrections to the Z-boson and the muon 
decay are contained in $a^\BZ$ and $a^\BW$, respectively, which is for
example reflected by the appearance of the corresponding three-point 
functions $C_1$, $C_2$, $C_3$ in $a^\BZ$ (see
\refapp{aux}). Of course, if we had used finite external lepton-masses 
$a^\BZ$ and $a^\BW$ would have also contained the complete 
dependence on these masses.

\btab \bce
$\begin{array}{|c|c|c||c|c|c|}
\hline
\Mt/\GeV &
\Delta x_\fer^\uni & \Delta x_\fer^\uni(dom) &
\MH/\GeV &
\Delta x_\bos^\uni & \Delta x_\bos^\uni(dom)
\\ \hline
100 & 3.21\times10^{-3} & 3.90\times10^{-3} &
100 & 2.17\times10^{-3} & 2.93\times10^{-3}
\\ \hline
140 &  7.50\times10^{-3} & 7.80\times10^{-3} &
300 &  1.39\times10^{-3} & 1.61\times10^{-3}
\\ \hline
 220 & 17.92\times10^{-3} & 18.04\times10^{-3} &
1000 &  0.12\times10^{-3} &  0.16\times10^{-3}
\\ \hline
\hline
\Mt/\GeV &
\Delta y_\fer^\uni & \Delta y_\fer^\uni(dom) &
\MH/\GeV &
\Delta y_\bos^\uni & \Delta y_\bos^\uni(dom)
\\ \hline
100 & -8.27\times10^{-3} & -7.69\times10^{-3} &
100 & -0.26\times10^{-3} & -0.37\times10^{-3}
\\ \hline
140 & -7.05\times10^{-3} & -6.80\times10^{-3} &
300 & -0.35\times10^{-3} & -0.37\times10^{-3}
\\ \hline
 220 & -5.68\times10^{-3} & -5.59\times10^{-3} &
1000 & -0.37\times10^{-3} & -0.37\times10^{-3}
\\ \hline
\hline
\Mt/\GeV &
\varepsilon_\fer^\uni & \varepsilon_\fer^\uni(dom) &
\MH/\GeV &
\varepsilon_\bos^\uni & \varepsilon_\bos^\uni(dom)
\\ \hline
100 & -6.25\times10^{-3} & -6.35\times10^{-3} &
100 & -2.80\times10^{-3} & -3.17\times10^{-3}
\\ \hline
140 & -6.00\times10^{-3} & -6.05\times10^{-3} &
300 & -3.59\times10^{-3} & -3.66\times10^{-3}
\\ \hline
 220 & -5.63\times10^{-3} & -5.64\times10^{-3} &
1000 & -4.19\times10^{-3} & -4.20\times10^{-3}
\\ \hline
\earr$
\caption{Comparison of the approximations, $a_{\fer/\bos}^\uni(dom)$,
for the universal parts of $\Delta x$, $\Delta y$, $\varepsilon$
with the exact values. The latter are obtained according to
(\protect\ref{domrem}) by evaluating the remainder terms,
$a_{\fer/\bos}^\uni (rem)$, given in \protect\refapp{xyerem}.}
\label{approx}
\ece
\etab
By inspection of \refta{approx} we get an impression of the quality
of the approximations $a_{\fer/\bos}^\uni \approx
a_{\fer/\bos}^\uni (dom)$. It turns out that the differences between
$a_{\fer/\bos}^\uni (dom)$ and $a_{\fer/\bos}^\uni$ are small
compared with (present and future) experimental errors, even for relatively
low values of $\Mt$ and $\MH$. More specifically, the differences are
$\lsim 0.5\times10^{-3}$, except for very low values of $\Mt,\MH\sim 100\GeV$.
Recall that the approximations are constructed such that they approach
the exact values asymptotically for $\Mt,\MH\gg\MZ$.

In order to explicitly display the magnitude of the different
contributions to
$\Delta x$, $\Delta y$, $\varepsilon$, we also list the corresponding
numerical expressions (based on $c_0 = 0.8768$):
\beqar
\Delta x_\fer^\uni (dom) &=&
(\;+0.52+1.34\log(t)+2.61\,t\;)\times10^{-3}, \nn\\
\Delta y_\fer^\uni (dom) &=&
(\;-7.94-1.34\log(t)\phantom{2.6+1\,t}\;)\times10^{-3}, \nn\\
\varepsilon_\fer^\uni (dom) &=&
(\;-6.43-0.45\log(t)\phantom{2.6+1\,t}\;)\times10^{-3},
\label{xyefernum}
\\[.5em]
\Delta x_\bos^\uni (dom) &=&
(\;+3.04-0.60\log(h)          \;)\times10^{-3}, \nn\\
\Delta y_\bos^\uni (dom) &=&
(\;-0.37\phantom{1.3+4\log(h)}\;)\times10^{-3}, \nn\\
\varepsilon_\bos^\uni (dom) &=&
(\;-3.13-0.22\log(h)          \;)\times10^{-3}, 
\label{xyebosnum}
\\[.5em]
\Delta x^\BZ &=& 0.09\times10^{-3}, \nn\\
\Delta y^\BZ &=& 8.43\times10^{-3}, \qquad
\Delta y^\BW  =  5.46\times10^{-3},  \nn\\
\varepsilon^\BZ &=& 4.06\times10^{-3},
\label{xyePDnum}
\eeqar
Note that $\Delta x^\BW$ and $\varepsilon^\BW$ vanish according to
(\ref{xewpd}).

As mentioned, the process dependent parts of $\Delta x, \Delta y$,
$\varepsilon$ were extracted by means of the pinch technique \cite{pt}.
Other procedures, previously used, lead to slightly different
decompositions, e.g.\
\beqar
\left.\Delta y^\BW\right\vert_{\text\citere{gauginv1}} \equiv
\delta_{vertex} + \delta_{box} &=&
\frac{\alpz}{4\pi s_0^2}\left[
6+\frac{2s_0^4-6s_0^2+7}{2s_0^2}\log(c_0^2)\right] \nn\\
&=& 7.34 \times 10^{-3}, \\[.5em]
\left.\Delta y^\BW\right\vert_{\text\citere{gauginv2}} \equiv
\varepsilon_{vertex} + \varepsilon_{box} &=&
\frac{\alpz}{4\pi s_0^2}\left[
6s_0^2-\frac{2s_0^4-10s_0^2+5}{2s_0^2}\log(c_0^2)\right] \nn\\
&=& 7.95 \times 10^{-3}, \hspace{2em}
\eeqar
of the (unique value of the) sum in (\ref{xyesplit}).

It seems appropriate to add a brief remark on the basic parameters in
our Lagrangians (\ref{cclag}) and (\ref{nclag2}) at this point. The
parameter $\Delta x$ according to (\ref{dx}) describes global $SU(2)$
violation in the ratio of the charged, $\MWpm$, and (unmixed) neutral
boson mass, $\MWO$. Accordingly, $\Delta x$ should be a process
independent quantity in the $SU(2)_I\times U(1)_Y$ theory. This is
indeed fulfilled, as according to (\ref{xyePDnum}) $\Delta x^\BZ\approx 0$.
Similarly, process independence must hold for the mass ratio
$\MWO^2/\MZ^2$. From the expression for the Z-boson mass in 
(\ref{nclag2}), with (\ref{swrel}) and (\ref{dy}), one obtains
\beqar
z = \frac{\MWO^2}{\MZ^2} &=& 1 - \frac{e^2(\MZ^2)}{g^2_\PWpm(0)}
\left(1+\Delta y-2\varepsilon\right) \nn\\
&=& 1 - \frac{e^2(\MZ^2)}{g^2_\PWpm(0)}
\Biggl(1+\Delta y^\uni_\fer +\Delta y^\uni_\bos
-2\varepsilon^\uni_\fer-2\varepsilon^\uni_\bos
\nn\\ && \phantom{1 - \frac{e^2}{g^2_\PWpm(0)}\Biggl(1}
+\Delta y^\BW+\frac{c_0^2}{s_0^2}\Delta x^\BZ\Biggr),
\label{dz}
\eeqar
where the last formula in (\ref{dxyepd}) was used in the second step.
For clarity, we added
the scale $g^2_\PWpm\equiv g^2_\PWpm(0)$ as an argument, as $g^2_\PWpm$ in
(\ref{cclag}) according to (\ref{gmu}) refers to the scale relevant for
muon decay. Noting that according to the definition (\ref{dy}),
$\Delta y^\BW$ removes the muon-decay
process-dependent part from $g^2_\PWpm(0)$, and neglecting the small
contribution due to $\Delta x^\BZ\approx 0.1\times 10^{-3}$, we indeed
find that also the mass ratio $z$ is a universal quantity at
one-loop level in the standard theory. We note that (\ref{dz})
may be rewritten in terms of our canonical input parameter $c_0$ as
\beq
z = c_0^2 + \frac{s_0^2c_0^2}{c_0^2-s_0^2}
\left(2\varepsilon+\Delta x-\Delta y\right).
\eeq
A final comment in this context concerns the significance of the last
relation in (\ref{dxyepd}) itself. Upon neglecting $\Delta x^\BZ$, it
reads $\varepsilon^\BZ\approx\Delta y^\BZ/2$. This is gratifying, as
finally only two independent process-dependent (Z- and W-vertex)
corrections remain, $\Delta y^\BZ$ and $\Delta y^\BW$. The appearance of
$\Delta y^\BZ$ as a linear contribution to the mixing parameter
$\varepsilon$, defined by (\ref{mixlag}) and (\ref{lameps}), is 
understood as a consequence of the fact that mixing effects can be 
represented by vertex modifications (and vice versa) via linear field 
redefinitions \cite{kn91}.

We have compared our numerical results with
the ones obtained by numerical evaluation based on the computer code
ZFITTER \cite{zfit} and on the formulae in \citere{nov93b}.
We found good agreement in general. We note, however, a small
discrepancy with the results obtained by evaluating the formulae
in \citere{nov93b}.
The discrepancy is due to the incorrect identification
\footnote{A correction of this point contained in \citere{novcor} was
brought to our attention upon circulating a preliminary version of the
present paper.}
of $\alpz$ (which according to its empirical evaluation \cite{alpz}
contains the
running of $\alpha (q^2)$ due to (light) fermion loops only)
in \citere{nov93b}
with the full expression for $\alpz$ which also contains (small)
effects from the top-quark and the $W$ bosons.

The fact that $\Delta y$ is practically independent of $\MH$ and
that it is the only parameter which depends on the process
specific corrections to muon decay (entering via the input parameter
$G_\mu$) underlines the
usefulness of our set of parameters for the analysis of the
experimental data.
Moreover recall that the explicit formulae for $\Delta x$, $\Delta y$,
$\varepsilon$ given here, together with formula (\ref{obslin}),
represent the complete one-loop results for the observables
$\MWpm/\MZ$, $\bar\sw^2$, $\Gamma_l$ within the electroweak standard
model, which have to be compared with the implicitly given and relatively
complex results existing in the literature.

We conclude our investigation of the parameters $\Delta x$, $\Delta y$,
$\varepsilon$ within the standard model with a short excursion to the
related decay $\PZ\to\nu_l\bar\nu_l$ $(l=\Pe,\mu,\tau)$. Making use of
the knowledge of $\Delta x$ and $\Delta y$, the decay width for
$\PZ\to\nu_l\bar\nu_l$, $\Gamma_\nu$, including all $\O(\alpha)$
corrections is given by
\beq
\Gamma_\nu = \Gamma_\nu^{(0)}
\left(1+ \Delta x - \Delta y + \delta_\nu^{rem}\right),
\eeq
with
\beqar
\Gamma_\nu^{(0)} &=& \frac{\alpz\MZ}{24s_0^2c_0^2}, \nn\\[.3em]
\delta_\nu^{rem} &=&
\frac{\alpz}{8\pi c_0^2}\left[
-48(1-2s_0^2)C_1 -8c_0^2(2-s_0^2)^2C_2
\right.\nn\\ &&\hphantom{\frac{\alpz}{8\pi c_0^2}}\left.
-4c_0^2(5-2s_0^2)\log(c_0^2)
-55+96s_0^2-8s_0^4
\right],
\eeqar
where $\delta_\nu^{rem}$ accounts for an additional (process
dependent) $SU(2)$-symmetry breaking, which, for simplicity, was not
incorporated into our effective Lagrangian. Note that all universal -- 
both bosonic and fermionic -- corrections to $\Gamma_\nu$ are included in
$\Delta x - \Delta y$, whereas $\delta_\nu^{rem}$ just corrects the ZPD
parts of $\Delta x - \Delta y$ by
\beq
\delta_\nu^{rem} = 2.86 \times 10^{-3}. 
\eeq

\section{Comparison with the experimental data}
\label{discuss}

In our comparison between theory and experiment, as in \citere{bi93},
we will proceed in two steps. In a first step, we will directly
compare the theoretical predictions (\ref{obslin}) with the empirical 
data in the three-dimensional space of $\MWpm/\MZ$, $\bar\sw^2$, 
$\Gamma_l$. In a second step, 
by using the inversion of (\ref{obslin}), which is explicitly given
in \citere{bi93}, we calculate the experimental values of the parameters
$\Delta x, \Delta y$, $\varepsilon$ from the data on $\MWpm/\MZ$, $\bar\sw^2$,
$\Gamma_l$ and compare them with our theoretical
predictions based on (\ref{xyesplit}) etc.

The relevant experimental data from the four LEP collaborations
and the CDF/UA2 value for the W mass are given by%
\footnote{These data are the most recent preliminary data presented at 
the La Thuile and Moriond conferences, March 1994 \cite{lep,sld}.}
\beqar
\MZ       &=& 91.1899\pm 0.0044\GeV, \nn\\
\MWpm/\MZ &=& 0.8814 \pm 0.0021,     \nn\\
\Gamma_l  &=& 83.98  \pm 0.18\MeV,   \nn\\
g_V/g_A \mbox{(all asymmetries LEP)} &=& 0.0711 \pm 0.0020. 
\eeqar
From the value of $g_V/g_A$ we derive
\beq
\bar\sw^2 \mbox{(all asymmetries LEP)} = 0.23223\pm 0.00050.
\label{swLEP}
\eeq
Inclusion of the preliminary value derived from the measurement of the
left-right asymmetry at the SLC by the SLD collaboration yields \cite{sld}
\beq
g_V/g_A \mbox{(all asymmetries LEP+SLD)} = 0.0737 \pm 0.0018,
\eeq
and consequently
\beq
\bar\sw^2 \mbox{(all asymmetries LEP+SLD)} = 0.23158\pm 0.00045.
\label{swLEPSLD}
\eeq
Both values (\ref{swLEP}) as well as (\ref{swLEPSLD}) for $\bar\sw^2$ 
will be confronted with our theoretical predictions.

The high precision of the experimental data is best appreciated by
comparing the data with the $\alpha(0)$ tree-level predictions based on
$\MZ$ and
\beqar
\alpha(0) &=& 1/137.0359895(61),  \nn\\
G_\mu     &=& 1.16639(2)10^{-5}\GeV^{-2},
\eeqar
as carried out e.g.\ in \citere{bi93}.
The figures shown there drastically demonstrate that the $\alpha(0)$
tree approximation is ruled out by several standard deviations.

Turning to a more refined analysis in \reffis{swmw}-\ref{mwgl}, we show 
the three projections of the 68\% C.L.\ volume
defined by the data in the three-dimensional $(\MWpm/\MZ, \bar\sw^2,
\Gamma_l)$-space in comparison with various theoretical predictions.
The point denoted by the symbol `star' corresponds to the
$\alpz$-tree-level prediction obtained from (\ref{obslin}) by putting
$\Delta x,\Delta y,\varepsilon\to 0$ and using 
$G_\mu$, $\MZ$, and \cite{alpz}
\beq
\alpz = 1/128.87\pm 0.12
\label{alpz}
\eeq
as input parameters.
While the $\alpz$-tree-level prediction only includes the
vacuum-polarization effects of the light fermions ($e^\pm, \mu^\pm,
\tau^\pm$ and the five light quarks) in the photon propagator (the `running'
$\alpha(q^2)$ between $q^2 = 0$ and $q^2=M^2_Z$),
a complete treatment of fermion-loop corrections also affects the
$\PWpm$ and the Z propagators and has to include the
top-quark loops. The corresponding predictions, obtained by inserting
$a_\fer^\uni$ $(a=\Delta x, \Delta y , \varepsilon)$
from (\ref{xyeferdom}) and \refapp{xyerem} into (\ref{obslin}),
are shown by the single lines%
\footnote{Note that the corresponding single lines in the figures of
\citere{bi93} were obtained by including only the top-quark
contributions.}
in \reffis{swmw}-\ref{mwgl}
for various values of $\Mt$. Finally, by inserting the complete 
one-loop expressions for $\Delta x$, $\Delta y$, $\varepsilon$
from (\ref{xyeferdom}) to (\ref{xewpd}) and \refapp{xyerem} into
(\ref{obslin}), one obtaines the three connected lines in 
\reffis{swmw}-\ref{mwgl}. They correspond to 
$\MH = 100, 300$ and 1000\GeV, respectively, while $\Mt$ varies from
$\Mt = 60\GeV$ to 240\GeV. Note that all theoretical predictions are
subject to the experimental uncertainty (\ref{alpz}) of the input
parameter $\alpz$. For the $\alpz$-tree-level prediction this
uncertainty is indicated by error bars in \reffis{swmw}-\ref{mwgl}.

We draw the following conclusions from \reffis{swmw}-\ref{mwgl}:
\renewcommand{\labelenumi}{(\roman{enumi})}
\begin{enumerate}
\item
The data have reached sufficient precision to allow one to start
the discrimination between the $\alpz$-tree-level, the
fermion-loop and the full one-loop prediction of the electroweak theory.
\item
The data start to establish a discrepancy from the fermion-loop
predictions, i.e.\ they start to require contributions beyond
the fermion-loop vacuum polarization to $\gamma$, Z and $\PWpm$
propagation.
\item
Such additional contributions are provided by the standard
`bosonic' effects which contain vertex corrections depending on
the trilinear boson self-couplings as well as vacuum-polarization
effects depending on trilinear and quadrilinear boson self-couplings
and the couplings of the Higgs scalar. In fact, the data
show a clear tendency to agree with the standard predictions, provided
a top quark exists in nature with a mass of the order of 150\GeV.
\end{enumerate}

In \reffis{dxe}-\ref{edy}, we show the experimental data for $\Delta x$,
$\Delta y$, $\varepsilon$ derived from the data on $M_W/M_Z$, 
$\bar\sw^2$, $\Gamma_l$ by employing the inversion of (\ref{obslin}) 
as given in \citere{bi93}. As in \reffis{swmw}-\ref{mwgl},
the data are compared with the
fermion-loop predictions and the full one-loop standard results.
Several interesting conclusions can immediately be drawn:
\begin{enumerate}
\item
First of all, the $SU(2)$ violating parameters $\Delta x$,
$\Delta y$, $\varepsilon$ in our basic Lagrangian 
(\ref{cclag}), (\ref{nclag2}) are
experimentally restricted to the magnitude of about $10 \times
10^{-3}$, which is the order of magnitude typically induced in these
parameters by standard one-loop electroweak corrections.
\item
As specifically seen in \reffi{dxe}, the fermion-loop predictions for
$\varepsilon$ and $\Delta x$ hardly differ from the full one-loop
results. In other words, the addition of the radiative effects originating
from standard bosonic couplings of the vector bosons among
each other and to the hypothetical Higgs scalar only leads to
minor effects in $\varepsilon$ and $\Delta x$. These cannot be resolved by
present (and future) precision data.
\item
\reffis{dxdy},\ref{edy}, however, reveal a spectacular difference 
between the fermion-loop and the full one-loop results. As inferred from
in \reffis{dxdy},\ref{edy},
the fermion-loop line of \reffi{dxe} lies far below the full one-loop
$\MH$-dependent lines in three-dimensional $(\varepsilon, \Delta x,
\Delta y)$-space. The shift in $\Delta y$ between the (uniquely 
determined) fermion-loop and the full one-loop results 
(by using the numerical formulae in (\ref{xyefernum}) to (\ref{xyePDnum}) )
is easily traced back to the sum of the process dependent corrections
$\Delta y^\BW + \Delta y^\BZ \approx 14 \times 10^{-3}$,
as the universal bosonic contribution, $\Delta y^\uni_\bos$, 
is negligibly small. The variable $\Delta y$ thus appears
to be the only one of our three parameters which is sensitive to
standard bosonic effects.
Moreover, these bosonic corrections are due to the sum of
process specific vertex and box corrections to muon decay
(entering via $G_\mu$) and the process specific vertex corrections to
Z decay into charged leptons. In other words, the
neutral-current data at the Z resonance start to 
quantitatively isolate the standard radiative corrections to the
charged-current muon-decay reaction in conjunction with the
vertex corrections to Z decay. 
\end{enumerate}

In summary, the deviations between the data and the fermion-loop
predictions in \reffis{swmw}-\ref{mwgl} find their origin in the fairly
substantial $\PWpm$- and Z-vertex corrections quantified by $\Delta y$.

\section{Conclusions}
\label{concl}

The empirical data on the $\MWpm$ mass and the leptonic
observables, $\bar\sw^2$ and $\Gamma_l$, at the Z mass, are
analysed in terms of an effective Lagrangian allowing for $SU(2)$
breaking via the parameters $\Delta x$, $\Delta y$ and $\varepsilon$.
Systematically discriminating between (trivial) fermion-loop
corrections to $\gamma$, Z and $\PWpm$ propagation and full
one-loop results we have given compact and simple analytic expressions
for $\Delta x$, $\Delta y$, $\varepsilon$, and consequently for
$\MWpm/\MZ$, $\bar\sw^2,$ $\Gamma_l$, in the standard $SU(2)_I\times 
U(1)_Y$ electroweak theory.

A comparison of the theoretical predictions with the most recent
data from LEP, SLD and CDF/UA2 shows that $SU(2)$ breaking
quantified by the parameters $\Delta x$, $\Delta y$, $\varepsilon$ is
restricted to the order of magnitude of $10 \cdot 10^{-3}$
typical for the order of magnitude of (standard) radiative
one-loop corrections. This fact in itself constitutes a major
triumph of the standard electroweak theory, as dramatic effects,
previously speculated upon, such as effects owing to e.g.\ near-by
compositeness, excited vector bosons etc., are excluded at a very high
level of precision.

Moreover, the data have become sufficiently accurate to start
to require genuine additional (virtual) effects beyond the 
$\alpz$-tree-level and the full fermion-loop predictions.
Such additional contributions are inherently contained in the
standard theory. The difference between the fermion-loop and 
the full one-loop results is specifically traced back to
important vertex corrections at the $\PWpm f\bar f'$ and the
$\PZ f\bar f$ vertices. This is revealed by
analysing the data in terms of the observable parameters
$\Delta x$, $\Delta y$, $\varepsilon$ derived from $\MWpm /\MZ$,
$\bar\sw^2$, $\Gamma_l$.

While present data are in excellent agreement with the standard
theory and even start to test fine details of the virtual
bosonic effects, the eventual direct verification of the
boson self-couplings and the discovery of the Higgs scalar
(or of whatever it stands for), apart from discovering the
top quark, seem indispensable for completely revealing the
structure of the electroweak phenomena in nature.

\appendix
\section*{Appendix}

\section{Remainders of the universal parts of $\Delta x$, $\Delta y$,
$\varepsilon$}
\label{xyerem}

In \refse{xyeps} formula (\ref{xyesplit}), we have defined the quantity 
$a_{\fer/\bos}^\uni(rem)$
as the difference between the exact universal part $a_{\fer/\bos}^\uni$
and the corresponding asymptotic limits for a heavy top quark 
($t\to\infty$) and a
heavy Higgs boson ($h\to\infty$). Although the contributions of these
`remainders' can be neglected for most practical purposes (compare
\refta{approx}), for completeness we give their analytic expressions.
For the fermionic contributions we obtain
\beqar
\Delta x_\fer^\uni (rem) &=&
\frac{\alpz}{8\pi s_0^2c_0^2}\left[
-\frac{1}{2}+\frac{3s_0^2}{2}-\frac{232s_0^4}{27}+\frac{256s_0^6}{27}
-\left(\frac{3}{2}+s_0^2+\frac{272s_0^4}{9}-\frac{320s_0^6}{9}\right)t
\right.\nn\\ &&\phantom{\frac{\alpz}{24\pi s_0^2}}\left.
\frac{t^2}{c_0^2}
+(t-c_0^2)^2(t+2c_0^2)\frac{1}{c_0^4}\log\left(1-\frac{c_0^2}{t}\right)
+\frac{\logx(t)}{1-4t}\left\{
c_0^2\left(1-\frac{32s_0^4}{9}\right)
\right.
\right.\nn\\ &&\phantom{\frac{\alpz}{24\pi s_0^2}}\left.
\left.
-\left(5-2s_0^2-\frac{64s_0^4c_0^2}{9}\right)t
+\left(4+2s_0^2+\frac{544s_0^4}{9}-\frac{640s_0^6}{9}\right)t^2
\right\}
\right],\nn\\[.3em]
\Delta y_\fer^\uni (rem) &=&
\frac{\alpz}{8\pi s_0^2}\left[
\frac{t^2}{c_0^4}
+\left(\frac{3}{2}+15s_0^2-\frac{464s_0^4}{9}+\frac{320s_0^6}{9}\right)
\frac{t}{c_0^2}
-\frac{3}{2}+\frac{8s_0^2}{3}-\frac{256s_0^4}{27}
\right.\nn\\ &&\phantom{\frac{\alpz}{24\pi s_0^2}}\left.
+\left(2-\frac{3t}{c_0^2}+\frac{t^3}{c_0^6}\right)
\log\left(1-\frac{c_0^2}{t}\right)
+\left\{(1-2t)\left(1-\frac{32s_0^4}{9}\right)
\right.
\right.\nn\\ &&\phantom{\frac{\alpz}{24\pi s_0^2}}\left.
\left.
-2t^2\left(1+16s_0^2-\frac{320s_0^4}{9}\right)
\right\}\frac{\logx(t)}{1-4t}
\right],\nn\\[.3em]
\varepsilon_\fer^\uni (rem) &=&
\frac{\alpz}{8\pi s_0^2}\left[
-\frac{19}{18}+\frac{232s_0^2}{27}-\frac{256s_0^4}{27}
-(39-272s_0^2+320s_0^4)\frac{t}{9}
+\left\{(1-2t)
\right.
\right.\nn\\ &&\phantom{\frac{\alpz}{24\pi s_0^2}}\left.
\left.
\times(-3+32s_0^2c_0^2)+2t^2(39-272s_0^2+320s_0^4)\right\}
\frac{\logx(t)}{9(1-4t)}
\right],
\label{xyeferrem}
\eeqar
and the bosonic contributions read
\beqar
\Delta x_\bos^\uni (rem) &=&
\frac{\alpz}{8\pi s_0^2c_0^2}\left[
\frac{47s_0^2}{12}-s_0^2h
+(1-2s_0^2)\frac{s_0^2h^2}{6c_0^2}
+\left(\frac{5c_0^2}{3}-\frac{3h}{2}+\frac{h^2}{2c_0^2}
-\frac{h^3}{12c_0^4}\right)\log(c_0^2)
\right.\nn\\ &&\phantom{\frac{\alpz}{24\pi s_0^2}}\left.
+s_0^2\left(\frac{5}{2}-\frac{3h}{2}
+(1-3s_0^2)\frac{h^2}{4c_0^2}
+(3-2s_0^2)\frac{s_0^2h^3}{12c_0^4}\right)\log(h)
\right.\nn\\ &&\phantom{\frac{\alpz}{24\pi s_0^2}}\left.
+\left(4+3s_0^2-(7+8s_0^2)\frac{h}{3}
+(8+13s_0^2)\frac{h^2}{12}-(1+2s_0^2)\frac{h^3}{12}\right)
\right.\nn\\ &&\phantom{\frac{\alpz}{24\pi s_0^2}}\left.
\times\frac{h}{h-4}\logy\left(\frac{1}{h}\right)
+\left(1-\frac{h}{3c_0^2}+\frac{h^2}{12c_0^4}\right)h
\logy\left(\frac{c_0^2}{h}\right)
\right],\nn\\[.3em]
\Delta y_\bos^\uni (rem) &=&
\frac{\alpz}{8\pi s_0^2}\left[
-\frac{47}{12}+(7-4s_0^2)\frac{h}{4c_0^2}
-\left(\frac{5}{6}-\frac{3h}{2}+\frac{3h^2}{4}-\frac{h^3}{6}\right)
\log(c_0^2)
\right.\nn\\ &&\phantom{\frac{\alpz}{24\pi s_0^2}}\left.
-(3-4s_0^2+2s_0^4)\frac{h^2}{6c_0^4}
+\left(c_0^2+(1+2c_0^6)\frac{h^4}{12c_0^6}-(11-3s_0^2)\frac{h}{2}
\right.
\right.\nn\\ &&\phantom{\frac{\alpz}{24\pi s_0^2}}\left.
\left.
+(17-12s_0^2+3s_0^4)\frac{h^2}{4c_0^2}
-(18-24s_0^2+15s_0^4-2s_0^6)\frac{h^3}{12c_0^4}\right)
\right.\nn\\ &&\phantom{\frac{\alpz}{24\pi s_0^2}}\left.
\times\frac{1}{h-c_0^2}\log\left(\frac{h}{c_0^2}\right)
-\left(3-\frac{8h}{3}+\frac{13h^2}{12}-\frac{h^3}{6}\right)
\frac{h}{h-4}\logy\left(\frac{1}{h}\right)
\right.\nn\\ &&\phantom{\frac{\alpz}{24\pi s_0^2}}\left.
+\left(1-\frac{h}{3c_0^2}+\frac{h^2}{12c_0^4}\right)\frac{h}{c_0^2}
\logy\left(\frac{c_0^2}{h}\right)
\right],\nn\\[.3em]
\varepsilon_\bos^\uni (rem) &=&
\frac{\alpz}{8\pi s_0^2}\left[
-\frac{67}{32}+h-\frac{h^2}{3}
-\left(\frac{5}{6}-\frac{3h}{2}+\frac{3h^2}{4}-\frac{h^3}{6}\right)\log(h)
\right.\nn\\ &&\phantom{\frac{\alpz}{24\pi s_0^2}}\left.
-\left(3-\frac{8h}{3}+\frac{13h^2}{12}-\frac{h^3}{6}\right)
\frac{h}{h-4}\logy\left(\frac{1}{h}\right)
\right].
\eeqar
By definition, $a^\uni_{\fer/\bos}(rem)$ approach asymptotically zero
with $t,h\to\infty$.
In other words, all powers $t^n$, $h^n$ $(n\ge0)$ cancel exactly in these
asymptotic expansions.

\section{Auxiliary functions}
\label{aux}

Here we list the explicit expressions for the auxiliary functions
which have been used in \refse{xyeps}. The functions $\logx$ and $\logy$
are given by
\beqar
&& \logx(x) = \Re\left[\beta_x\log\left(\frac{\beta_x-1}
              {\beta_x+1}\right)\right], \quad\mbox{with}\;\;
             \beta_x=\sqrt{1-4x+i\epsilon}, \nn\\
&& \logy(x) = \Re\left[\beta_x^*\log\left(\frac{1-\beta_x^*}
              {1+\beta_x^*}\right)\right], \nn\\[.3em]
\mbox{more explicitly,} \hspace*{3em} && \nn\\
0<x<\frac{1}{4}: && \logx(x) = \logy(x) =
\sqrt{1-4x}\log\left(\frac{1-\sqrt{1-4x}}{1+\sqrt{1-4x}}\right), \nn\\
x>\frac{1}{4}:
&& \logx(x) = \sqrt{4x-1}\left(2\atn\sqrt{4x-1}-\pi\right), \nn\\
&& \logy(x) = 2\sqrt{4x-1}\atn\sqrt{4x-1}.
\eeqar
The constants $C_1, C_2, C_3$ are shorthands for the scalar three-point
integrals occuring in the process dependent parts of the Z decay,
\beqar
C_1 &=& \MZ^2\Re\left[C_0(0,0,\MZ^2,0,\MZ,0)\right] =
-\frac{\pi^2}{12} = -0.8225, \nn\\[.3em]
C_2 &=& \MZ^2\Re\left[C_0(0,0,\MZ^2,0,\MW,0)\right] =
\frac{\pi^2}{6}-\Re\left[\Li\left(1+\frac{1}{c_0^2}\right)\right]
= -0.8037, \nn\\[.3em]
C_3 &=& \MZ^2\Re\left[C_0(0,0,\MZ^2,\MW,0,\MW)\right] \nn\\
&=& \Re\left[\log^2\left(\frac{i\sqrt{4c_0^2-1}-1}
                          {i\sqrt{4c_0^2-1}+1}\right)\right]
= -\left(\pi-2\atn\sqrt{4c_0^2-1}\right)^2 = -1.473.
\eeqar
The first three arguments of the $C_0$ function label the external momenta
squared, the last three the inner masses of the corresponding vertex
diagram.

\clearpage
\begin{figure}
\begin{center}
\begin{picture}(16,16)
\put(0,-5){\includegraphics{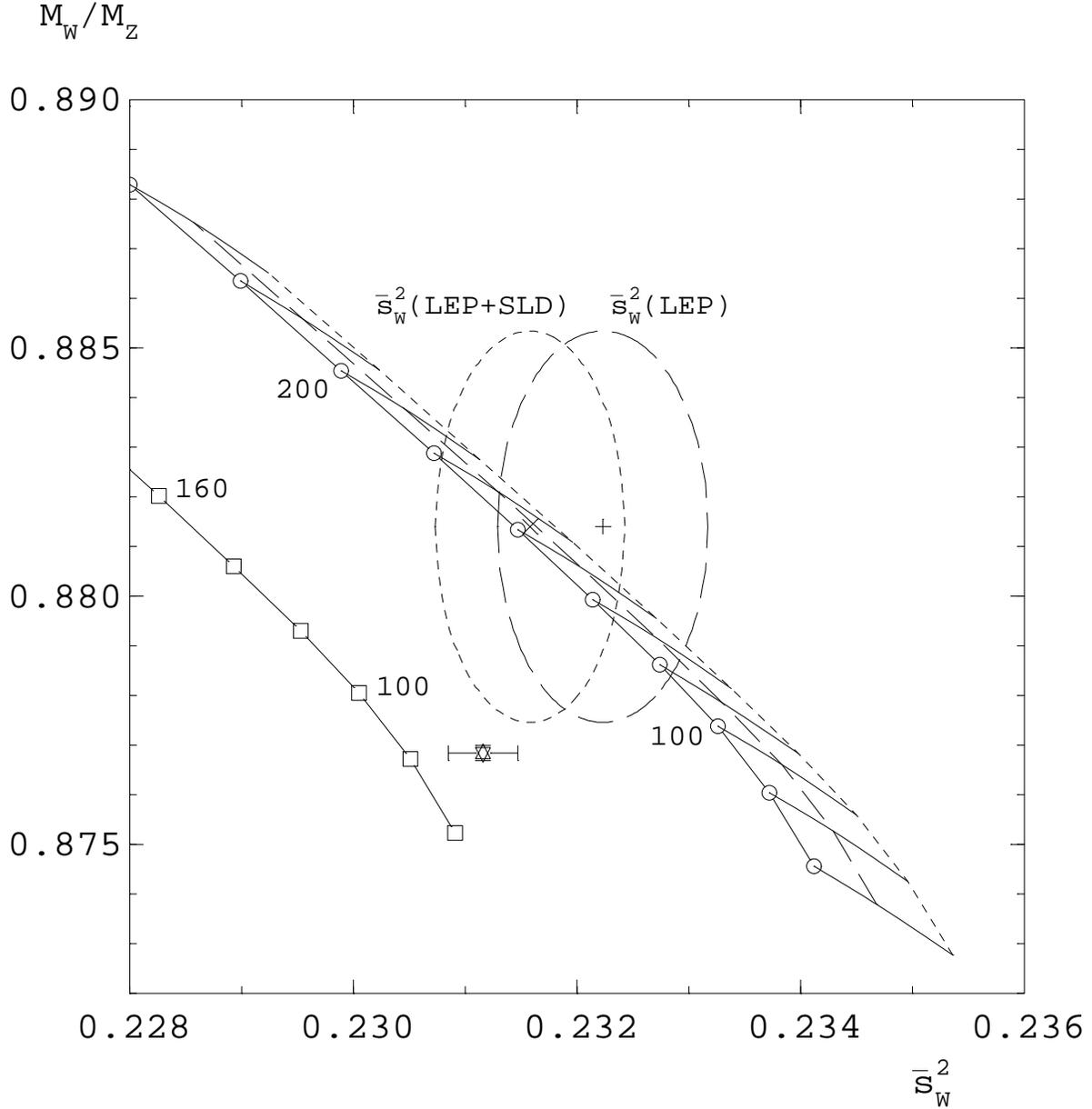}}
\end{picture}
\end{center}
\caption{The experimental data are represented by the projection of the
68\% C.L.\ volume in $(\MWpm/\MZ, \bar\sw^2, \Gamma_l)$-space into the
$(\MWpm/\MZ, \bar\sw^2)$-plane. The symbol `star' denotes the
$\alpz$-tree-level prediction. The fermion-loop prediction is shown by
the single line, the squares indicating steps in 20\GeV\ in $\Mt$. The
full one-loop standard-model prediction is shown for Higgs-boson masses
of $\MH=100\GeV$ (solid line), 300\GeV\ (long-dashed line), 1000\GeV\ 
(short-dashed line), the circles indicating steps in 20\GeV\ in $\Mt$.}
\label{swmw}
\end{figure}

\clearpage
\begin{figure}
\begin{center}
\begin{picture}(16,16)
\put(0,-5){\includegraphics{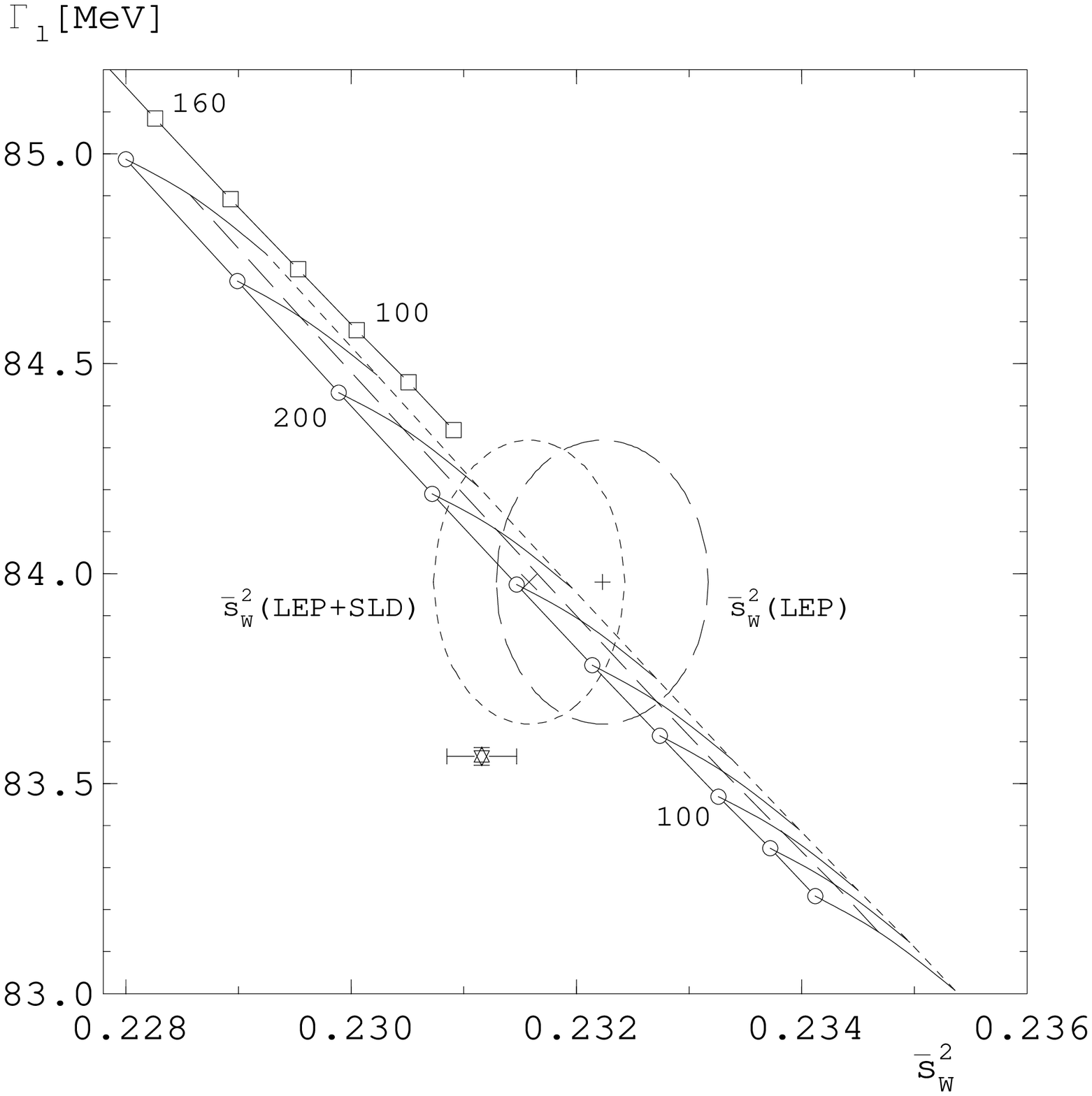}}
\end{picture}
\end{center}
\caption{Same signature as \protect\reffi{swmw}, but for the 
$(\bar\sw^2, \Gamma_l)$-plane.}
\label{swgl}
\end{figure}

\clearpage
\begin{figure}
\begin{center}
\begin{picture}(16,16)
\put(0,-5){\includegraphics{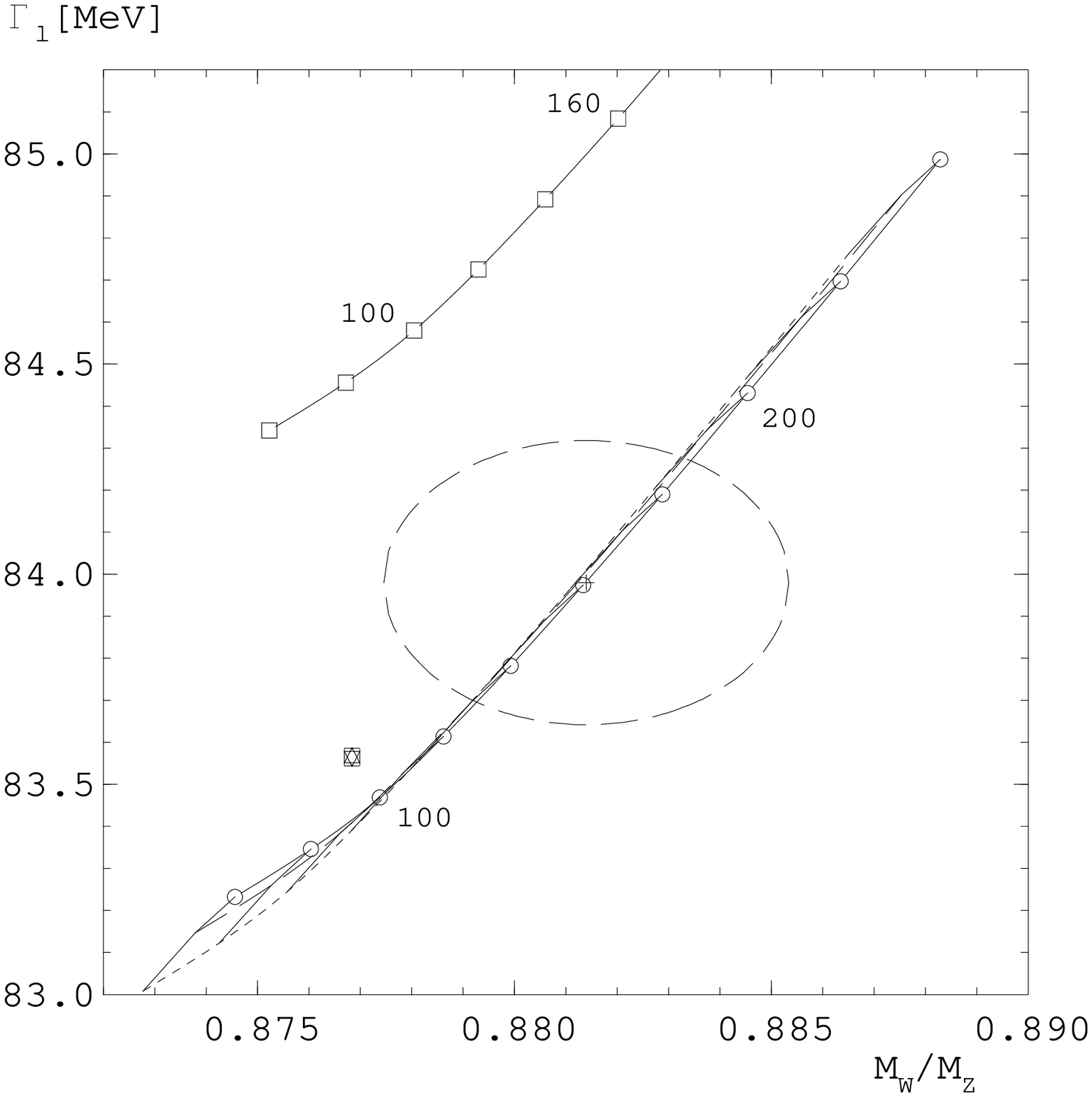}}
\end{picture}
\end{center}
\caption{Same signature as \protect\reffi{swmw}, but for the 
$(\MWpm/\MZ, \Gamma_l)$-plane.}
\label{mwgl}
\end{figure}

\clearpage
\begin{figure}
\begin{center}
\begin{picture}(16,16)
\put(0,-5){\includegraphics{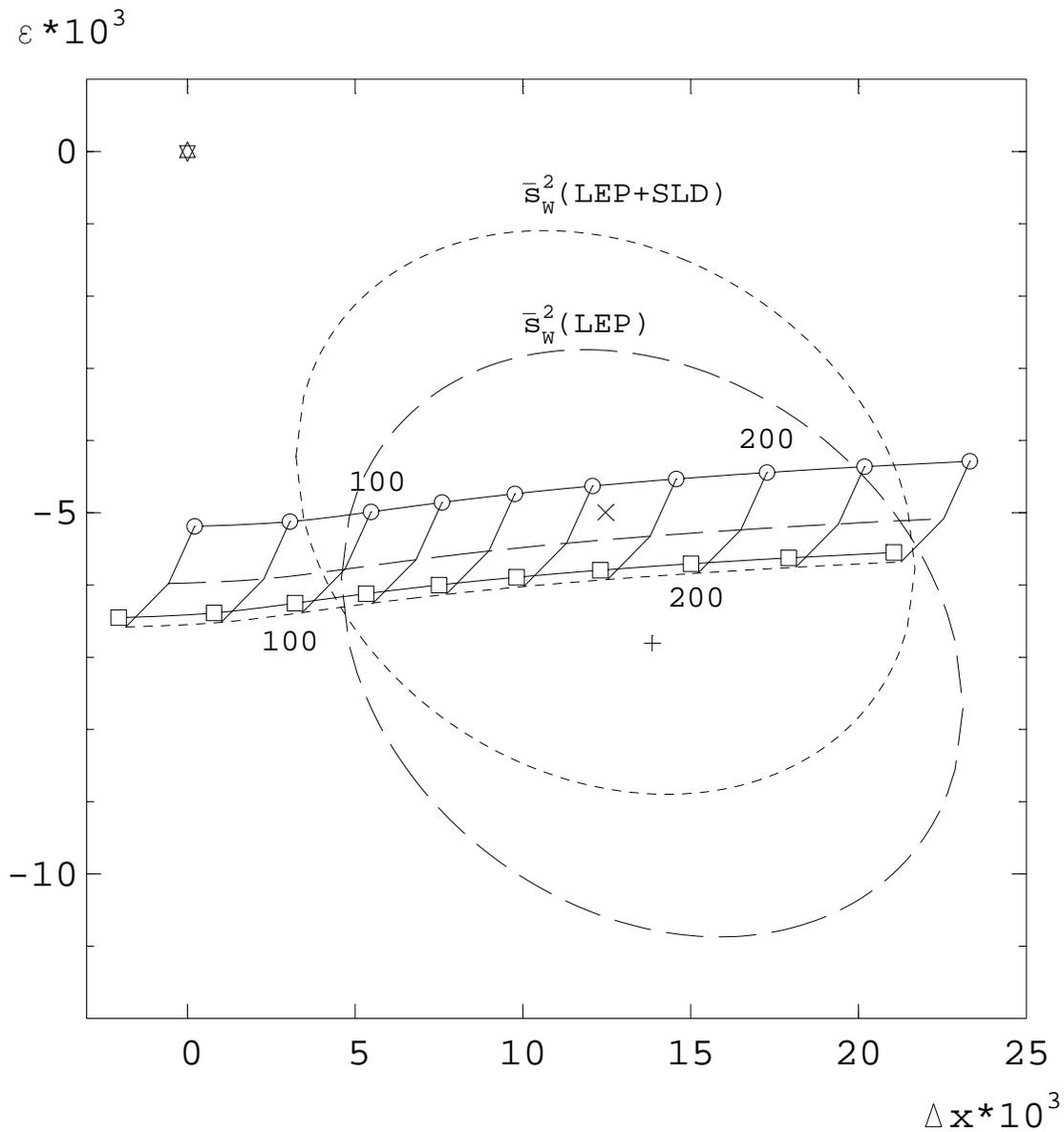}}
\end{picture}
\end{center}
\caption{The experimental data are represented by the projection of the
68\% C.L.\ volume in $(\varepsilon, \Delta x, \Delta y)$-space into the
$(\varepsilon, \Delta x)$-plane. As in 
\protect\reffis{swmw}-\protect\ref{mwgl}, 
the fermion-loop prediction is shown by the single line, the
full one-loop standard-model prediction by the connected lines.}
\label{dxe}
\end{figure}

\clearpage
\begin{figure}
\begin{center}
\begin{picture}(16,16)
\put(0,-5){\includegraphics{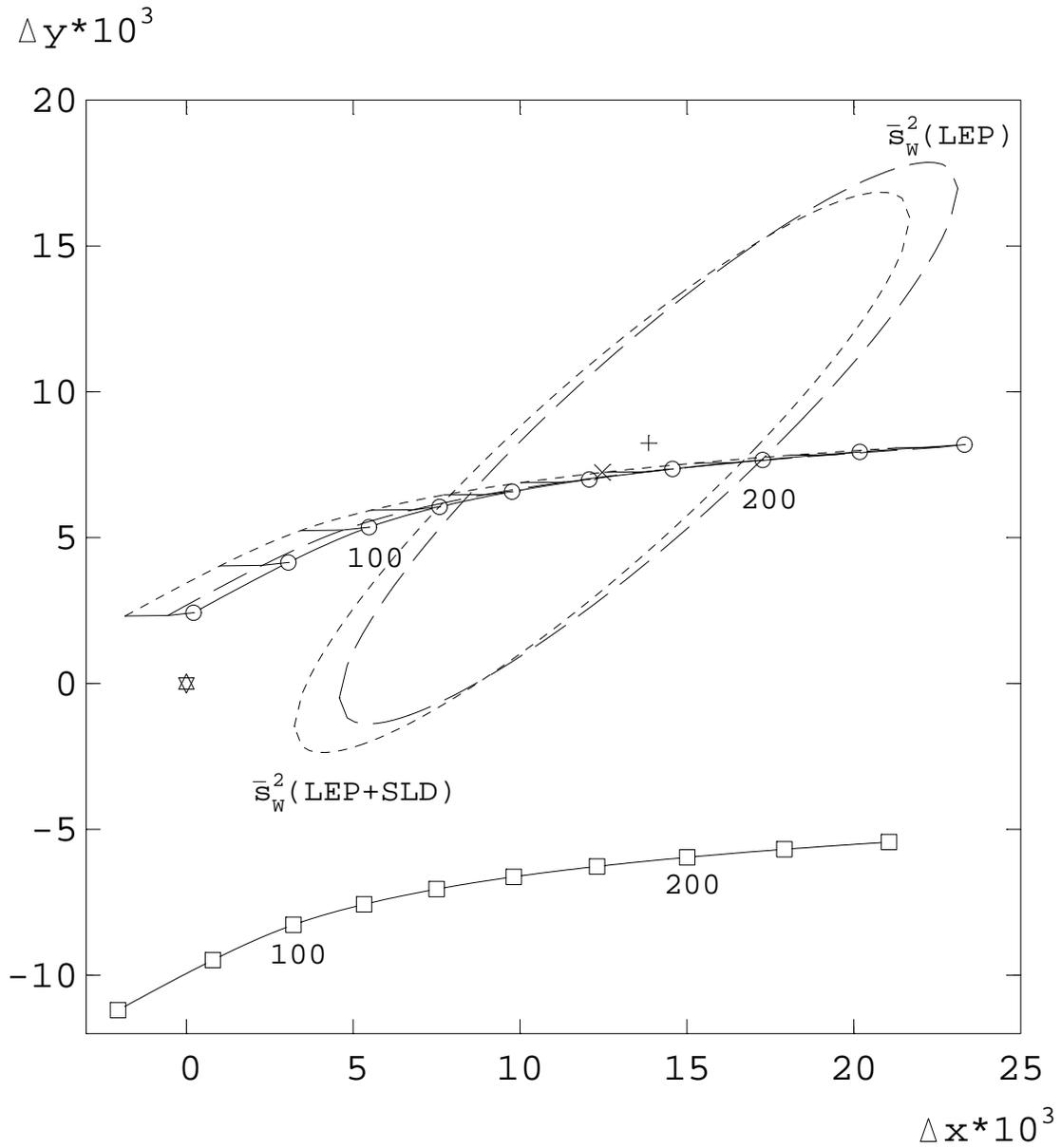}}
\end{picture}
\end{center}
\caption{Same signature as \protect\reffi{dxe}, but for the parameters
$\Delta x$ and $\Delta y$.}
\label{dxdy}
\end{figure}

\clearpage
\begin{figure}
\begin{center}
\begin{picture}(16,16)
\put(0,-5){\includegraphics{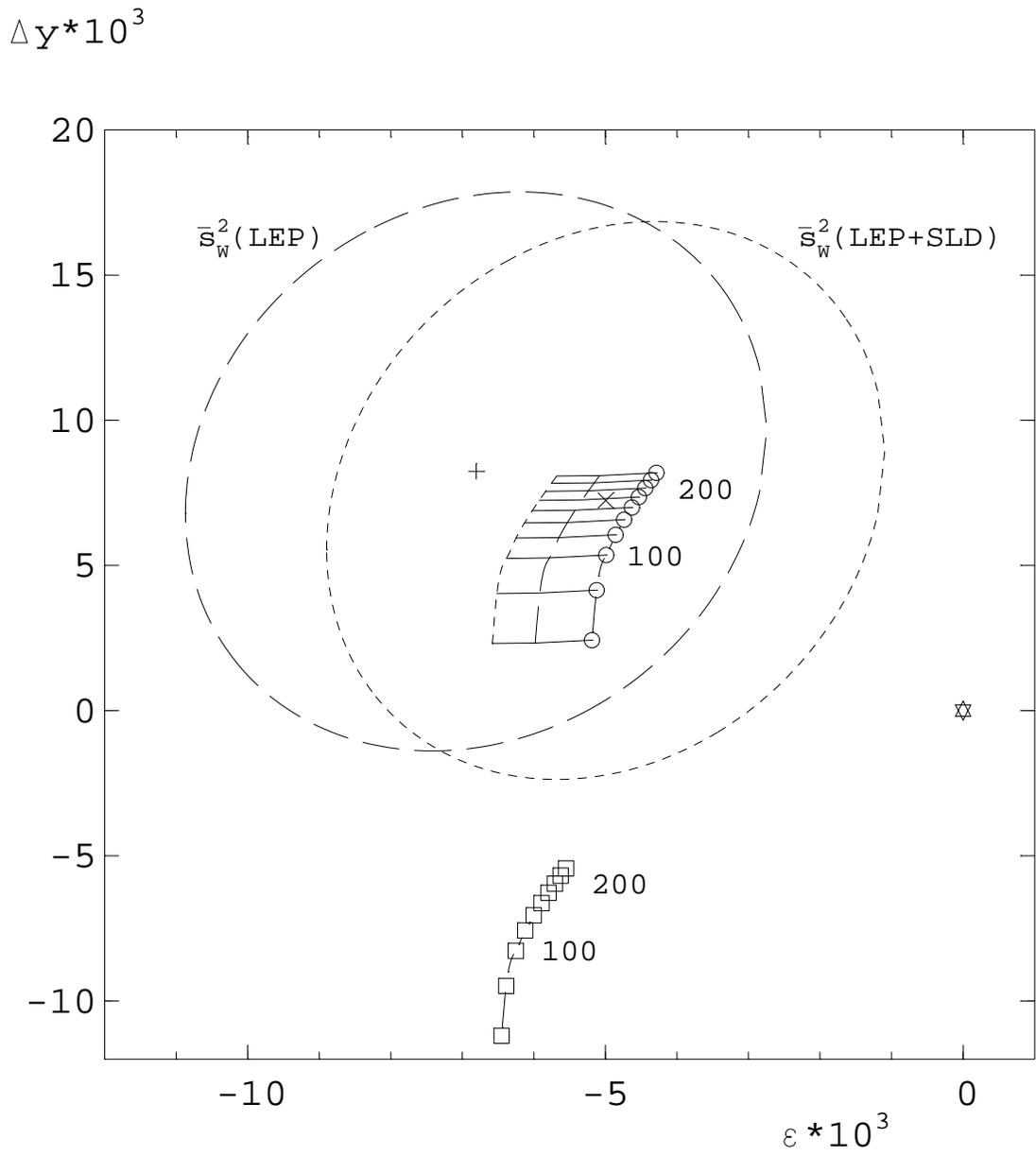}}
\end{picture}
\end{center}
\caption{Same signature as \protect\reffi{dxe}, but for the parameters
$\varepsilon$ and $\Delta y$.}
\label{edy}
\end{figure}


\begin{thebibliography}{99}
\frenchspacing
\newcommand{\zpC}[3]{{\sl Z. Phys.} {\bf C#1} (19#2) #3}
\newcommand{\zp}[3]{{\sl Z. Phys.} {\bf #1} (19#2) #3}
\newcommand{\npB}[3]{{\sl Nucl. Phys.} {\bf B#1} (19#2) #3}
\newcommand{\plB}[3]{{\sl Phys. Lett.} {\bf B#1} (19#2) #3}
\newcommand{\prD}[3]{{\sl Phys. Rev.} {\bf D#1} (19#2) #3}
\newcommand{\pr}[3]{{\sl Phys. Rev.} {\bf #1} (19#2) #3}
\newcommand{\prl}[3]{{\sl Phys. Rev. Lett.} {\bf #1} (19#2) #3}
\newcommand{\rmp}[3]{{\sl Rev. Mod. Phys.} {\bf #1} (19#2) #3}
\newcommand{\fp}[3]{{\sl Fortschr. Phys.} {\bf #1} (19#2) #3}
\newcommand{\cpc}[3]{{\sl Comp. Phys. Comm.} {\bf #1} (19#2) #3}
\newcommand{\nim}[3]{{\sl Nucl. Instr. Meth.} {\bf #1} (19#2) #3}
\newcommand{\ej}[3]{{\bf #1} (19#2) #3}
\newcommand{\jcp}[3]{{\sl J. Comp. Phys. } {\bf #1} (19#2) #3}

\bibitem{bi93} M.\ Bilenky, K.\ Kolodziej, M.\ Kuroda, and D.\
Schildknecht, \plB{319}{93}{319}; \\
D.\ Schildknecht, {\sl Proceedings of the International Conference on
High Energy Physics}, Marseille, July 1993.

\bibitem{go88} G.\ Gounaris and D.\ Schildknecht, \zpC{40}{88}{447},
\zpC{42}{89}{107}.

\bibitem{kn91} J.-L.\ Kneur, M.\ Kuroda, and D.\ Schildknecht,
\plB{262}{91}{93},

\bibitem{hu78} P.Q.\ Hung and J.J.\ Sakurai, \npB{143}{78}{81}; \\
J.D.\ Bjorken, \prD{19}{79}{335}.

\bibitem{alpz} H.\ Burkhardt, F.\ Jegerlehner, G.\ Penso, and C.\
Verzegnassi, \zpC{43}{89}{497}.

\bibitem{gsw} S.L.\ Glashow, \npB{22}{61}{579}; \\
S.\ Weinberg, \prl{19}{67}{1264}; \\
A.\ Salam, in {\it Elementary Particle Theory}, ed.\ N.\ Svartholm
(Almqvist and Wiksell, Stockholm, 1968), p.\ 367.

\bibitem{al93} G.\ Altarelli, R.\ Barbieri, and F.\ Caravaglios,
\npB{405}{93}{3};\\
G.\ Altarelli, CERN-TH 6867/93, {\it Schladming lecture}, February 
1993.

\bibitem{admex} A.\ Denner,
\fp{41}{93}{307}; \\
M.\ B\"ohm and A.\ Denner, {\sl Radiative Corrections in the Electroweak
Standard Model}, in {\sl Proceedings of the Workshop on High
Energy Phenomenology}, Mexico City, 1991,
eds. R. Huerta and M.A. Perez (World Scientific, Singapore, 1992), p.~1.

\bibitem{gauginv1} D.C.\ Kennedy, \npB{321}{89}{83}; \\
D.C.\ Kennedy and B.W.\ Lynn, \npB{322}{89}{1}; \\
B.W.\ Lynn, Stanford University Report SU-ITP-867 (1989); \\
D.C.\ Kennedy, FNAL Report FERMI-CONF-91/271-T (1992).

\bibitem{gauginv2} M.\ Kuroda, G.\ Moultaka, and D.\ Schildknecht,
\npB{350}{91}{25}.

\bibitem{pt} 
J.M.\ Cornwall, \prD{26}{82}{1453}; \\
J.M.\ Cornwall and J.\ Papavassiliou, \prD{40}{89}{3474}; \\
J.\ Papavassiliou, \prD{41}{90}{3179}; \\
G.\ Degrassi and A.\ Sirlin, \prD{46}{92}{3104}.

\bibitem{ve77} M.\ Veltman, \npB{123}{77}{89}.

\bibitem{zfit} D.\ Bardin et al., CERN-TH.\ 6443/92.

\bibitem{nov93b} V.A.\ Novikov, L.B.\ Okun, and M.I.\ Vysotsky,
\npB{397}{93}{35}.

\bibitem{novcor} V.A.\ Novikov, L.B.\ Okun, and M.I.\ Vysotsky,
CERN-TH.\ 7071/93.

\bibitem{lep} LEP collaborations, {\it La Thuile and Moriond
Conferences}, March 1994.

\bibitem{sld} SLD collaboration, {\it La Thuile and Moriond
Conferences}, March 1994.

\end{thebibliography}
\end{document}